\documentclass[12pt,reqno]{article}

\global\arraycolsep=1pt

\usepackage{amsmath,cite,amsfonts}
\usepackage{array}
\usepackage{graphicx,caption}
\usepackage{float}
\usepackage{multicol}
\usepackage[english]{babel}
\usepackage{hyperref}

\usepackage{mathrsfs}
\usepackage[Symbol]{upgreek}
\usepackage{bbm}
\usepackage{dsfont}
\usepackage{amssymb}
\usepackage{wasysym}

\newcommand{\be}{\begin{equation}}
\newcommand{\ee}{\end{equation}} 
\newcommand{\bea}{\begin{eqnarray}} 
\newcommand{\eea}{\end{eqnarray}}

\let\a=\alpha \let\b=\beta   \let\e=\epsilon
\let\z=\zeta \let\h=\eta   
  \let\n=\nu \let\x=\xi  
   \let\f=\phi  \let\y=\psi
\let\w=\omega      \let\G=\Gamma   
 \let\P=\Pi    
\let\C=\Chi 

\def\nn{\nonumber} \def\bd{\begin{document}} \def\ed{\end{document}}
\def\ds{\documentstyle} \let\fr=\frac \let\bl=\bigl \let\br=\bigr
\let\Br=\Bigr \let\Bl=\Bigl
\let\bm=\bibitem
\let\na=\nabla
\let\pa=\partial \let\ov=\overline

\def\ba{\begin{array}}
\def\ea{\end{array}}

\def\del{\partial}
\def\vp{\varphi}
\def\st#1{{\scriptstyle #1}}
\def\sst#1{{\scriptscriptstyle #1}}
\def\oneone{\rlap 1\mkern4mu{\rm l}}
\def\td{\tilde}
\def\wtd{\widetilde}

\def\dalemb#1#2{{\vbox{\hrule height .#2pt
        \hbox{\vrule width.#2pt height#1pt \kern#1pt
                \vrule width.#2pt}
        \hrule height.#2pt}}}
\def\square{\mathord{\dalemb{6.8}{7}\hbox{\hskip1pt}}}

\newcommand{\ra}{\rightarrow}
\newcommand{\lra}{\longrightarrow}
\newcommand{\Lra}{\Leftrightarrow}
\newcommand{\ap}{\alpha^\prime}
\newcommand{\bp}{\tilde \beta^\prime}
\newcommand{\tr}{{\rm tr} }
\newcommand{\Tr}{{\rm Tr} }

\newcommand{\og}{\overline{g}}
\newcommand{\oB}{\overline{B}}
\newcommand{\ophi}{\overline{\phi}}
\newcommand{\oF}{\overline{F}}
\newcommand{\oD}{\overline{\nabla}}
\newcommand{\oC}{\overline{\Gamma}}
\newcommand{\bI}{\mathbb{I}}
\newcommand{\cA}{\mathcal{A}}
\newcommand{\cI}{\mathcal{I}}
\newcommand{\cL}{\mathcal{L}}
\newcommand{\cO}{\mathcal{O}}
\newcommand{\fX}{\mathfrak{X}}
\newcommand{\fY}{\mathfrak{Y}}

\newenvironment{subitemize}{\begin{itemize}\scriptsize}{\end{itemize}}

\def\ie{{\it i.e.\ }}
\def\eg{{\it e.g.\ }}
\def\eg,{{\it e.g.,}}
\def\viz{{\it viz.}\ }
\def\nn{\nonumber}

\def\ft#1#2{\tfrac{#1}{#2}}
\def\fft#1#2{\frac{#1}{#2}}

\def\0{{\sst{(0)}}}
\def\1{{\sst{(1)}}}
\def\2{{\sst{(2)}}}
\def\3{{\sst{(3)}}}
\def\4{{\sst{(4)}}}
\def\5{{\sst{(5)}}}
\def\6{{\sst{(6)}}}
\def\7{{\sst{(7)}}}
\def\8{{\sst{(8)}}}
\def\n{{\sst{(n)}}}
\def\cA{{{\cal A}}}
\def\cF{{{\cal F}}}
\def\tV{\widetilde V}
\def\tW{\widetilde W}
\def\tH{\widetilde H}
\def\tE{\widetilde E}
\def\tF{\widetilde F}
\def\tA{\widetilde A}
\def\im{{{\rm i}}}
\def\tY{{{\wtd Y}}}
\def\ep{{\epsilon}}
\def\vep{{\varepsilon}}

\def\bD{{{\bar D}}}
\def\sech{\text{sech}}

\newfont{\bbbold}{msbm10}
\newfont{\sbbbold}{msbm10 scaled 800}
\def\bbA{\mbox{\bbbold A}}
\def\bbB{\mbox{\bbbold B}}
\def\bbC{\mbox{\bbbold C}}
\def\bbD{\mbox{\bbbold D}}
\def\bbE{\mbox{\bbbold E}}
\def\bbF{\mbox{\bbbold F}}
\def\bbG{\mbox{\bbbold G}}
\def\bbH{\mbox{\bbbold H}}
\def\bbI{\mbox{\bbbold I}}
\def\bbJ{\mbox{\bbbold J}}
\def\bbK{\mbox{\bbbold K}}
\def\bbL{\mbox{\bbbold L}}
\def\bbM{\mbox{\bbbold M}}
\def\bbN{\mbox{\bbbold N}}
\def\bbO{\mbox{\bbbold O}}
\def\bbP{\mbox{\bbbold P}}
\def\bbQ{\mbox{\bbbold Q}}
\def\bbR{\mbox{\bbbold R}}
\def\bbS{\mbox{\bbbold S}}
\def\bbT{\mbox{\bbbold T}}
\def\bbU{\mbox{\bbbold U}}
\def\bbV{\mbox{\bbbold V}}
\def\bbW{\mbox{\bbbold W}}
\def\bbX{\mbox{\bbbold X}}
\def\bbY{\mbox{\bbbold Y}}
\def\bbX{\mbox{\bbbold X}}
\def\bbZ{\mbox{\bbbold Z}}

\def\C{{\mathscr{C}}}
\def\N{{\mathcal{N}}}
\def\M{{\mathcal{M}}}
\def\V{{\mathcal{V}}}
\def\w{{\scriptstyle W}}

\def\G{{\mathfrak{G}}}
\def\H{{\mathfrak{H}}}
\def\P{{\mathfrak{P}}}
\def\I{{\mathfrak{I}}}
\def\J{{\mathfrak{J}}}
\def\j{{\mathfrak{j}}}

\def\Z{\mathds{Z}}

\def\e{{\boldsymbol{e}}}
\def\f{{\boldsymbol{f}}}
\def\h{{\boldsymbol{h}}}
\def\x{{\boldsymbol{x}}}
\def\y{{\boldsymbol{y}}}
\def\b{{\boldsymbol{b}}}
\def\z{{\boldsymbol{z}}}
\def\upsi{{{\boldsymbol \upsilon}}}

\newcommand{\Scal}[1]{\Bigl ({#1} \Bigr )}
\newcommand{\scal}[1]{\bigl ({#1} \bigr )}
\newcommand{\trace}{\hbox {Tr}~}

\newcommand{\imperial}{\it The Blackett Laboratory, Imperial College London\\
Prince Consort Road, London SW7 2AZ}

\newcommand{\auth}{
K.S. Stelle\footnote{email: k.stelle@imperial.ac.uk}}

\begin{document}

\begin{center}  

{\Large {\bf Mass gaps and braneworlds}} \\
\bigskip
In Memory of Peter Freund

\vspace{15pt}

\auth

\vspace{7pt}
{\imperial} 

\vspace{30pt}

\end{center}  
%%%%%%%%%%%%%%%%%%%
\abstract{
Remembering the foundational contributions of Peter Freund to supergravity, and especially to the problems of dimensional compactification, reduction is considered with a non-compact space transverse to the lower dimensional theory. The known problem of a continuum of Kaluza-Klein states is avoided here by the occurrence of a mass gap between a single normalizable zero-eigenvalue transverse wavefunction and the edge of the transverse state continuum. This style of reduction does not yield a formally consistent truncation to the lower dimensional theory, so developing the lower-dimensional effective theory requires integrating out the Kaluza-Klein states lying above the mass gap.
}
%%%%%%%%%%%%%%%%%%%
\begin{center}
\rule{50pt}{1pt}
\end{center}
\vspace{.3cm}

\section*{\large Memories of Peter Freund}

It is with great fondness that I think back to all the various interactions that I had with Peter Freund throughout my career. Of course, there are the many shared interests in physics, especially in supersymmetry, nonabelian gauge theories of all sorts, dimensional reduction and string theory. But there are also the episodes, and especially the story telling about episodes, at which Peter was a world master. One could not say that Peter was generally softly spoken. One of my earliest memories of Peter was at an Institute for Theoretical Physics workshop at the University of California at Santa Barbara back in 1986. Peter was giving a seminar, and, as usual, electronic amplification was hardly needed for him. However, one of our senior colleagues (who shall remain nameless) was sitting in the front row and was actually managing to sleep during Peter's seminar. This was in the original UCSB Institute, on the top floor of Ellison Hall. Achieving sleep during one of Peter's seminars provoked a certain amount of amused commentary amongst the audience. However, at one point during the seminar a characteristically Californian event took place: an earthquake! And being at the top of the building, the motion was clearly felt. What did the audience do -- run out? No: the main reaction was to lean forward and see if even an earthquake wasn't enough to disturb a slumberer able to sleep during one of Peter's forcefully presented seminars.

\section*{\large Peter, Dimensional Reduction and other Enduring Topics}

The fact that supergravity and superstring theories originate most naturally in higher spacetime dimensions -- 11 for maximal supergravity and 10 for superstring theories -- gave rise to intensive research on reduction schemes starting in the early 1980s. A key achievement was made in the 1980 paper by Peter Freund and Mark Rubin on the reduction via an $S^7$ transverse geometry from $D=11$ down to $D=4$ spacetime dimensions \cite{Freund:1980xh}. In this highly influential paper, the ``ground state'' maximally symmetric geometry in $D=4$ proved to be an Anti de Sitter space. The reduction mechanism involved turning on flux for the 4-form antisymmetric-tensor field strength of the $D=11$ theory, as well as a warped-product structure for the overall higher dimensional spacetime. All of these features have remained prominent in the subsequent development of string and supergravity theories: the key roles of warped products, Anti de Sitter vacua and the importance of flux vacua.

In related work, Peter explored cosmological dimensional reduction schemes in which the effective dimensionality of spacetime is not maximally symmetric but time dependent \cite{Freund:1982pg}. Then, in a paper together with Phillial Oh \cite{Freund:1985je}, Peter attacked the thorny problem of reduction from $D=10$, $N=1$ supergravity plus Yang-Mills down to $D=4$, for which a ``no-go'' theorem had been claimed \cite{Freedman:1983zt}. The metric of the $D=4$ spacetime was again not maximally symmetric. At that time, before recognition of the r\^ole that could be played by Calabi-Yau reduction spaces, the focus was mainly on sphere and toroidal reductions. All of Peter's 1980s contributions have continued to be greatly influential up to the present in the continuing effort to understand the cosmological implications of supergravity and superstring theories.

Of course, there is much more to learn from Peter's large volume of original research. There is in particular the importance of topology in quantum gravity and the Higgs mechanism \cite{Arafune:1974uy,Eguchi:1976db} -- a topic whose central importance is now widely recognized. Another of Peter's topics which intertwines with much of current research was the characterization of gauge fields as Nambu-Goldstone fields for the nonlinear realization of a higher symmetry \cite{Cho:1975sf}.

\section*{\large The Universe as a Membrane}

In contrast to dimensional reduction schemes on compact spaces, another possibility might be that a lower dimensional spacetime is embedded into a higher dimensional spacetime with a noncompact transverse space. The idea of formulating the cosmology of our universe on a brane embedded in a higher-dimensional spacetime dates back, at least, to Rubakov and Shaposhnikov \cite{Rubakov:1983bb}. Attempts in a supergravity context to achieve a localization of gravity on a brane embedded in an infinite transverse space were made by Randall and Sundrum \cite{Randall:1999vf} and by Karch and Randall \cite{Karch:2000ct} using patched-together sections of $\hbox{AdS}_5$ space with a delta-function source at the joining surface. This produced a ``volcano potential'' for the effective Schr\"odinger problem in the direction transverse to the brane, giving rise to a bound state concentrating gravity in the 4D directions.

\begin{figure}
\centering
\includegraphics[scale=.6]{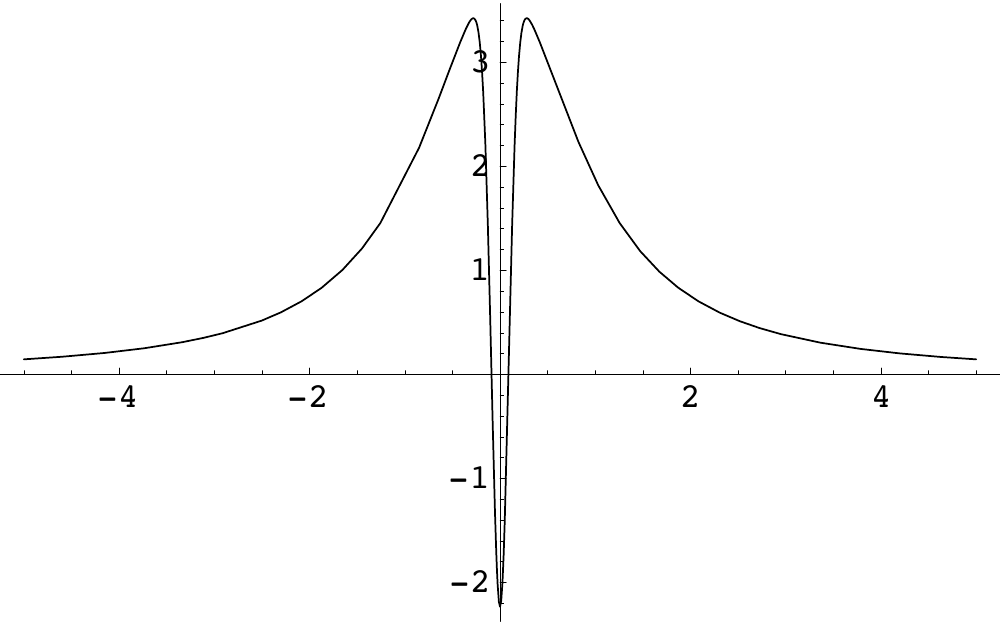}
\caption{Volcano potential}
\end{figure}

Attempting to embed such models into a full supergravity/string-theory context proved to be problematic, however. Splicing together sections of $\hbox{AdS}_5$ is clearly an artificial construction that does not make use of the natural D-brane or NS-brane objects of string or supergravity theory.
\bigskip

These difficulties were studied more generally by Csaki, Erlich, Hollowood and Shirman \cite{Csaki:2000fc} and then by Bachas and Estes \cite{Bachas:2011xa}, who traced the difficulty in obtaining localization within a string or supergravity context to the behavior of the warp factor for the 4D subspace. In the Karch-Randall spliced model, one obtains a peak in the warp factor at the junction:
\begin{figure}[H]
\centering
\includegraphics[scale=.45]{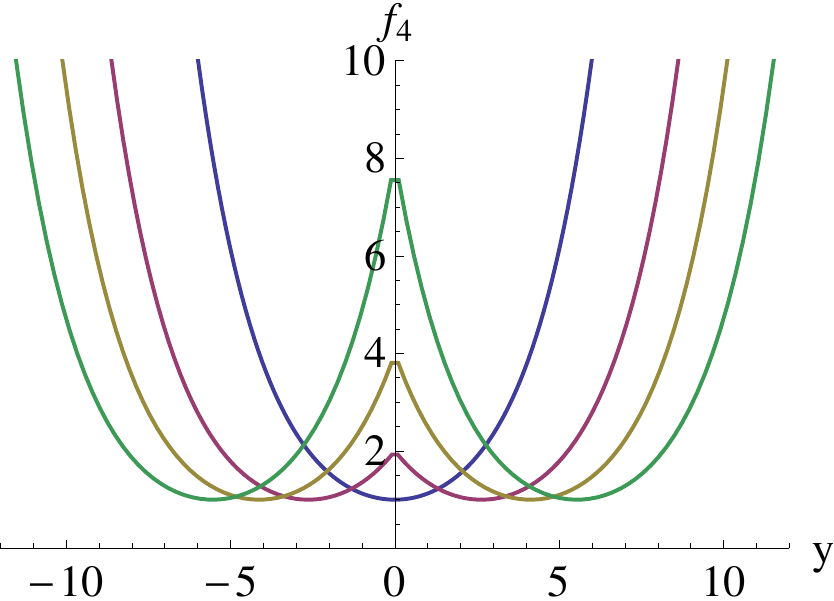}
\caption{Peaked warp factor}
\end{figure}
However, in a string or supergravity context, warp factors tend to join smoothly, even in a ``Janus'' type construction \cite{Bachas:2011xa}.

Here's why Bachas and Estes considered that one could not have a natural localization of gravity on a brane with an infinite transverse space. Consider fluctuations away from a smooth $D$-dim background 
\be 
d\hat s^2=e^{2A(z)}(\eta_{\mu\nu}+h_{\mu\nu}(x)\xi(z))dx^\mu dx^\nu + \hat g_{ab}(z)dz^adz^b\,,
\ee where $\xi(z)$ is the transverse wave function. Such a transverse wave function with eigenvalue $\lambda$ needs to satisfy the transverse wave equation
\be
{e^{-2A}\over\sqrt{\hat g}}\partial_a(\sqrt{\hat g}e^{4A}\hat g^{ab}\partial_b)\xi=-\lambda\xi\,.
\ee
The norm of $\xi(z)$ is then given by \be \lambda||\xi||^2=-\int d^{D-4}z\xi(\partial_a\sqrt{\hat g}e^{4A}\hat g^{ab}\partial_b\xi)\,.\ee

If one {\em assumes} that one may integrate by parts without producing a surface term, then one would have $\lambda||\xi||^2\longrightarrow\int d^{D-4}z\sqrt{\hat g}e^{4A}|\partial\xi|^2$. Consequently, if one is looking for a transverse wavefunction $\xi$ with $\lambda=0$ as needed for massless 4D gravity, one would need to have $\partial_a\xi=0$ yielding $\xi={\rm constant}$, which is not normalizable in an infinite transverse space.

The resolution of this problem requires very specific self-adjointness features of the transverse wavefunction problem, to which we shall return.

\section*{\large Another Approach:\\ Salam-Sezgin Theory and its Embedding}

Abdus Salam and Ergin Sezgin constructed in 1984 a version of 6D minimal (chiral, \ie (1,0)) supergravity coupled to a 6D 2-form tensor multiplet and a 6D super-Maxwell multiplet which gauges the $U(1)$ R-symmetry of the theory \cite{Salam:1984cj}. This Einstein-tensor-Maxwell system has the bosonic Lagrangian
\bea
{\cal L}_{\mbox{\scriptsize SS}}&=&\frac12 R - \frac1{4g^2}e^\sigma F_{\mu\nu}F^{\mu\nu}-\frac 1{6}e^{-2\sigma}G_{\mu\nu\rho}G^{\mu\nu\rho}-\frac 12\partial_\mu\sigma\partial^\mu\sigma-g^2e^{-\sigma}\cr
G_{\mu\nu\rho} &=& 3\partial_{[\mu}B_{\nu\rho]} + 3 F_{[\mu\nu} A_{\rho]}\,.
\eea
Note the {\em positive} potential term for the scalar field $\sigma$. This is a key feature of all R-symmetry gauged models generalizing the Salam-Sezgin model, leading to models with noncompact symmetries. For example, upon coupling to yet more vector multiplets, the sigma-model target space can have a structure $SO(p,q)/(SO(p)\times SO(q))$.

The Salam-Sezgin theory does not admit a maximally symmetric 6D solution, but it does admit a $\hbox{(Minkowski)}_4\times S^2$ solution with the flux for a $U(1)$ monopole turned on in the $S^2$ directions
\bea
ds^2&=&\eta_{\mu\nu}dx^\mu dx^\nu + a^2(d\theta^2 + \sin^2\theta d\phi^2)\cr
A_m dy^m &=& (n/2g)(\cos\theta \mp 1)d\phi\cr
\sigma&=&\sigma_0=\hbox{const}\ ,\qquad B_{\mu\nu}=0\cr
g^2&=&{e^{\sigma_0}\over 2a^2}\ ,\qquad\qquad\ \  n=\pm1\,.
\eea
Requiring the flux quantum number to be $n=\pm1$ amounts to constructing the Hopf fibration of $S^3$.

A way to obtain the Salam-Sezgin theory from M theory was given by Cveti\v{c}, Gibbons and Pope \cite{Cvetic:2003xr}. This employed a reduction from 10D type IIA supergravity on the space ${\cal H}^{(2,2)}$, or, equivalently, from 11D supergravity on $S^1\times {\cal H}^{(2,2)}$. The ${\cal H}^{(2,2)}$ space is a cohomogeneity-one 3D hyperbolic space which can be obtained by embedding into $R^4$ via the condition $\mu_1^2+\mu_2^2-\mu_3^2-\mu_4^2=1$. 
This embedding condition is $\hbox{SO}(2,2)$ invariant, but the embedding $\bbR^4$ space just has $\hbox{SO}(4)$ symmetry, so the linearly realized isometries of this space are just \resizebox{5cm}{!}{$\hbox{SO}(2,2)\cap\hbox{SO}(4)=\hbox{SO}(2)\times\hbox{SO}(2)$}. The cohomogeneity-one ${\cal H}^{(2,2)}$ metric can be written
\be
ds_3^2 = \cosh 2\rho d\rho^2+\cosh^2\!\rho d\alpha^2+\sinh^2\!\rho d\beta^2\,.
\ee

Since ${\cal H}^{(2,2)}$ admits a natural $SO(2,2)$ group action, the resulting 7D supergravity theory has maximal (32 supercharge) supersymmetry and a gauged $SO(2,2)$ symmetry, linearly realized on $SO(2)\times SO(2)$. Note how this fits neatly into the general scheme of extended Salam-Sezgin gauged models.

The reduced $D=7$ bosonic Lagrangian is given by  \cite{Cvetic:2003xr}
\bea
{\cal L}_7 &=& R\, {*\oneone} - \ft5{16}\, \Phi^{-2}\, {*d\Phi}\wedge d\Phi - {\ast p_{\a \beta}}
\wedge p^{\a\beta} - \ft12 \Phi^{-1}\, 
{\ast H_\3}\wedge  H_\3 \nn\\
&& - \ft12 \Phi^{-1/2}\, 
\pi_{\bar A}{}^\a \, \pi_{\bar B}{}^{\beta}\, \pi_{\bar C}{}^\a\, 
\pi_{\bar D}{}^\beta\, 
{\ast F_\2^{\bar A\bar B}}\wedge F_\2^{\bar C \bar D}
 - \fft1{g} \,  \Omega -  V \, {*\oneone}\,,
\eea
where $\pi_{\bar A}{}^\a$ ($\st{\bar A}\  \&\  \st{\a} = 1\ldots4$) are scalar vielbeins describing\\ 9 = 16 - 1 (det) - 6 (gauge) degrees of freedom. 
\begin{itemize}
\item{The global ``composite'' group structure is revealed in the covariant derivative 
\be
p_{\a\beta} = {\pi^{-1}}_{(\a}{}^{\bar A}\,
[\delta_{\bar A}{}^{\bar B}\, d + g\, A_{\1\bar A}{}^{\bar B}]\, 
\pi_{\bar B}{}^\gamma\, \delta_{\beta)\gamma}\nn
\ee 
whose $\alpha$, $\beta$ indices are always raised and lowered with $\delta_{\alpha\beta}$, showing that there are no scalar ghosts.}
\item{The local gauge symmetry (gauge field $ A_{\1\bar A}{}^{\bar B}$) acts on the $\bar A$, $\bar B$ indices, preserving a metric $\eta_{\bar A \bar B}$}. If $\eta_{\bar A \bar B}=\hbox{diag}(++++)$, then one has the standard local $\hbox{SO}(4)$ symmetry \cite{Nastase:1999cb}. 
\item{If $\eta_{\bar A \bar B}=\hbox{diag}(++--)$, then one has a local $\hbox{SO}(2,2)$ symmetry.}
\item{The scalar field potential is given by
\be
V = \ft12 g^2\, \Phi^{1/2}\, ( 2 M_{\a\beta}\, M_{\a\beta} 
-(M_{\a\a})^2)\nn
\ee
built from the unimodular matrix
\be
M_{\a\beta} = {\pi^{-1}}_\a{}^{\bar A}\, 
{\pi^{-1}}_\beta{}^{\bar B}\,  \eta_{\bar A \bar B}\nn
\ee
}
\end{itemize}

The ${\cal H}^{(2,2)}$ reduced theory in 7D can be further truncated to minimal (16 supercharge) 7D supersymmetry, and then yet further reduced on $S^1/\Z_2$ to obtain precisely the $(1,0)$ 6D Salam-Sezgin gauged $U(1)$  supergravity theory. This naturally admits the $\hbox{(Minkowski)}_4\times S^2$ Salam-Sezgin ``ground state'' solution. Moreover, the result of this chain of reductions from 11D or 10D is a mathematically consistent truncation: every solution of the 6D Salam-Sezgin theory can be lifted to an exact solution in 10D type IIA or 11D supergravity. Such a consistent truncation needs to be made in the standard Kaluza-Klein fashion, however, suppressing all dependence on the ${\cal H}^{(2,2)}$ reduction space coordinates.

At the quantum level, the original Salam-Sezgin theory has a $U(1)$ anomaly. 
Instead of dimensionally reducing the 7D precursor theory on $S^1$, however, the 7th dimension can be $S^1/\Z_2$ compactified in a Ho\v{r}ava-Witten construction, producing naturally a chiral generalization of the Salam-Sezgin theory in 6 dimensions. 
Consideration of anomaly inflow together with coupling of appropriate boundary hypermultiplets and tensor multiplets then allows the construction of anomaly-free generalizations of the Salam-Sezgin model \cite{Pugh:2010ii}.
\section*{\large The Kaluza-Klein Spectrum}

Reduction on the non-compact ${\cal H}^{(2,2)}$ space from ten to seven dimensions, despite its mathematical consistency, does not provide a full physical basis for compactification to 4D. The chief problem is that the truncated Kaluza-Klein modes form a {\it continuum} instead of a discrete set with mass gaps. Moreover, the wavefunction of ``reduced'' 4D states when viewed from 10D or 11D in a standard Kaluza-Klein reduction includes a non-normalizable factor owing to the infinite ${\cal H}^{(2,2)}$ directions. Accordingly, the higher-dimensional supergravity theory does not naturally localize gravity in the lower-dimensional subspace when handled by ordinary Kaluza-Klein methods.

The $D=10$ lift of the Salam-Sezgin ``vacuum'' solution yields the metric
\bea
ds_{10}^2 &=& (\cosh2\rho)^{1/4}\, \Big[e^{-\ft14\bar\phi}\, d\bar s_6^2 +
  e^{\ft14\bar\phi}\, dy^2 + \ft12\bar g^{-2}\, 
   e^{\ft14\bar\phi}\, \Big(d\rho^2 \cr
&&+\ft14[d\psi+ 
   \sech2\rho(d\chi-2\bar g \bar A)]^2 + \ft14(\tanh2\rho)^2\,
  (d\chi-2\bar g \bar A)^2\Big)\Big]
\cr
 \bar A_\1&=& -\fft1{2\bar g}\, \cos\theta\, d\varphi
  \eea
  in which the $ d\bar s_6^2$ metric has $\hbox{Minkowski}_4\times S^2$ structure
  \be
  d\bar s_6^2 = dx^\mu dx^\nu\eta_{\mu\nu} + \fft1{8\bar g^2}\, 
 (d\theta^2+\sin^2\theta d\varphi^2)\,.
 \ee
 
 Instead of suppressing dependence on all of the reduction coordinates, we now adopt a different procedure in developing the lower dimensional effective theory. The inclusion of gravitational fluctuations about the above background may be accomplished by replacing
 \be
 \eta_{\mu\nu}\longrightarrow  \eta_{\mu\nu}+ h_{\mu\nu}(x,z)\,,
 \ee
 where the $z^p$ are reduction-space coordinates transverse to the 4D coordinates $x^\mu$.
 
 \section*{\large Bound States and Mass Gaps\footnote{This development was made in collaboration with Chris Pope and Ben Crampton \cite{Crampton:2014hia}.}}
 
 An approach to obtaining the localization of gravity on the 4D subspace is to look for a {\em normalizable} transverse-space wavefunction $\xi(z)$ for $h_{\mu\nu}(x,z)=h_{\mu\nu}(x)\xi(z)$ with a {\em mass gap} before the onset of the continuous massive Kaluza-Klein spectrum. This could be viewed as analogous to an effective field theory for electrons confined to a metal by a nonzero work function.

General study of the fluctuation spectra about brane solutions shows that the mass spectrum of spin-two fluctuations about a brane background is given by the spectrum of the scalar Laplacian in the transverse embedding space of the brane \cite{Csaki:2000fc,Bachas:2011xa}\bea
\square_{(10)}F &=& {1\over\sqrt{-\det g_{(10)}}}\partial_M\left(\sqrt{-\det g_{(10)}} g^{MN}_{(10)}\partial_N F\right)\cr
&=& H_{\mbox{\scriptsize SS}}^{\frac14}(\square_{(4)} + g^2 \triangle_{\theta,\phi,y,\psi,\chi} + g^2 \triangle_{\mbox{\scriptsize KK}})\cr
\hbox{warp factor:}\ &H_{\mbox{\scriptsize SS}}& = \sech 2\rho\,;\quad \triangle_{\mbox{\scriptsize KK}}={\partial^2\over\partial \rho^2}+{2\over\tanh(2\rho)}{\partial\over\partial\rho}\,.
\eea

The $z^p$ directions $\theta,\phi,y,\psi\ \&\ \chi$ are all compact, and for them one can employ ordinary Kaluza-Klein reduction methods, truncating to the invariant sector for these coordinates, but retaining dependence on the single noncompact coordinate $\rho$.

To handle the noncompact direction $\rho$, one needs to expand all fields in eigenmodes of $ \triangle_{\mbox{\scriptsize KK}}$:
\be
\phi(x^\mu,\rho)=\sum_i\phi_{\lambda_i}(x^\mu)\xi_{\lambda_i}(\rho) + \int_\Lambda^\infty d\lambda\phi_\lambda(x^\mu)\xi_\lambda(\rho)\,,
\ee
where the $\phi_{\lambda_i}$ are discrete eigenmodes and the $\phi_\lambda$ are the continuous Kaluza-Klein eigenmodes. Their eigenvalues give the Kaluza-Klein masses in 4D from $\square_{(10)}\phi_\lambda=0$ using $\triangle_{\theta,\phi,y,\psi,\chi} \phi_\lambda=0$ (with $g=\sqrt2 \bar g$ now)
\bea
 \triangle_{\mbox{\scriptsize KK}}\xi_\lambda&=&-\lambda\xi_\lambda\cr
\square_{(4)}\phi_\lambda&=&(g^2\lambda)\phi_\lambda\,.
\eea

One can rewrite the $\triangle_{\mbox{\scriptsize KK}}$ eigenvalue problem as a Schr\"odinger equation by making the substitution
\be
\Psi_\lambda(\rho)=\sqrt{\sinh(2\rho)}\,\xi_\lambda(\rho)\,,
\ee
after which the eigenfunction equation takes the Schr\"odinger equation form
\be
-{d^2\Psi_\lambda\over d\rho^2} + V(\rho)\Psi_\lambda=\lambda\Psi_\lambda
\ee
where the potential is
\be
V(\rho)=2-{1\over \tanh^2(2\rho)}\,.
\ee

The Salam-Sezgin Schr\"odinger equation potential $V(\rho)$ asymptotes to the value 1 for large $\rho$. In this limit, the Schr\"odinger equation becomes
\be
{d^2\Psi_\lambda\over d\rho^2}+4e^{-4\rho}\Psi_\lambda+(\lambda-1)\Psi_\lambda=0\,,
\ee
giving scattering-state solutions for $\lambda>1$:
\be
\Psi_\lambda(\rho)\sim \left(A_\lambda e^{i\sqrt{\lambda-1}\rho} + B_\lambda e^{-i\sqrt{\lambda-1}\rho} \right)\quad\text{ for large $\rho$}\,,
\ee
while for $\lambda<1$ one can have $L^2$ normalizable bound states. Recalling the $\rho$ dependence of the measure $\sqrt{-g_{(10)}}\sim(\cosh(2\rho))^{\frac14}\sinh(2\rho)$, one finds for large $\rho$
\be
\int_{\rho_1\gg1}^\infty \vert\Psi_\lambda(\rho)\vert^2d\rho<\infty
\Rightarrow \Psi_\lambda\sim B_\lambda e^{-\sqrt{1-\lambda}\rho} \ \mbox{for } \lambda<1\,,
\ee
so for $\lambda<1$ one has candidates for bound states.

\section*{\large Puzzles of the Schr\"odinger Problem and the Zero-mode Bound State}

The limit as $\rho\rightarrow0$ of the potential $V(\rho)=2-{1/\tanh^2(2\rho)}$ is just $V(\rho)=-1/(4\rho^2)$. The associated Schr\"odinger problem has a long history as one of the most puzzling cases in one-dimensional quantum mechanics. It has been studied and commented upon over the decades by Von Neumann; Pauli; Case; Landau \& Lifshitz; de Alfaro, Fubini \& Furlan; and many others. 

A key feature of this 1D problem is its $SO(1,2)$ {\em conformal invariance}. This symmetry has the consequence that, at the classical level, there is no way to form a definite scale for the transverse Laplacian eigenvalue of an $L^2$ normalizable ground state. (Except for the value zero, which is what will happen, as we shall see.) Discussion of the corresponding quantum theory requires a regularization that breaks this 1D conformal symmetry and gives rise to the choice of a {\em self-adjoint extension} for the domain of the Laplacian in order to determine the ground state. 

The $-\ft14$ coefficient is key to the peculiarity of this Schr\"odinger problem: for coefficients greater than $-\ft14$, there is no $L^2$ normalizable ground state, while for coefficients less than $-\ft14$, an infinity of $L^2$ normalizable discrete bound states appear. 
For the precise coefficient $-\ft14$, a regularized treatment shows the existence of a {\em single} $L^2$ normalizable bound state separated by a mass gap and lying below the continuum of scattering states \cite{Essin2005}. The precise eigenvalue of this ground state, however, is not fixed by normalizability considerations and hence remains, so far, a free parameter of the quantum theory.

Although the full $V(\rho)=2-{1/\tanh^2(2\rho)}$ potential breaks the 1D conformal invariance away from $\rho=0$, it nonetheless shares with the $V=-1/(4\rho^2)$ problem the indeterminacy of the ground-state eigenvalue. The Schr\"odinger potential $V(\rho)=2-\coth^2(2\rho)$ diverges as $\rho\rightarrow 0$; this is a regular singular point of the Schr\"odinger equation. Near $\rho=0$, solutions have a structure given by a Frobenius expansion
\be
\Psi_\lambda\sim\sqrt\rho(C_\lambda+D_\lambda\log\rho)\,.
\ee

This behavior at the origin does not affect $L^2$ normalizability, but it does indicate that we have a family of candidate bound states characterized by $\theta=\arctan({C_\lambda\over D_\lambda})$. 
Indeed, numerical study shows that there is a $1\leftrightarrow1$ relationship between $\theta$ and the eigenvalue $\lambda$. Moreover, the behavior of a candidate wavefunction $\xi_\lambda$ is logarithmic as $\rho\rightarrow 0$, in contrast to the non-singular character of the underlying Salam-Sezgin spacetime.

This 1-D quantum mechanical system with the $V(\rho)=2-\coth^2(2\rho)$ potential belongs to a special class of { P\"oschl-Teller} integrable systems. Neither normalizability nor self-adjointness are by themselves sufficient to completely determine the transverse wavefunction for the reduced effective theory, \ie the value of the parameter $\theta$. A key feature of such systems, however is 1-D {\em supersymmetry} and requiring that this be unbroken by the transverse wavefunction $\Psi_\lambda$ selects the value $\lambda=0$.

The self-adjointness condition requires selection of just one value of $\theta=\arctan({C_\lambda\over D_\lambda})$. The structure of general candidate $\Psi_\lambda$ eigenfunctions can't be given in terms of standard functions, but for $\lambda=0$, the Schr\"odinger equation can be solved exactly. The normalized result, corresponding to $\theta=0$, is
\be
\Psi_0(\rho)=\sqrt{\sinh(2\rho)}\,\xi_0(\rho)=\frac{2\sqrt{3}}{\pi}\sqrt{\sinh(2\rho)}\,\log(\tanh\rho)\,.
\ee

\begin{figure}[H]
\centering
\captionsetup{width=\linewidth}
\includegraphics{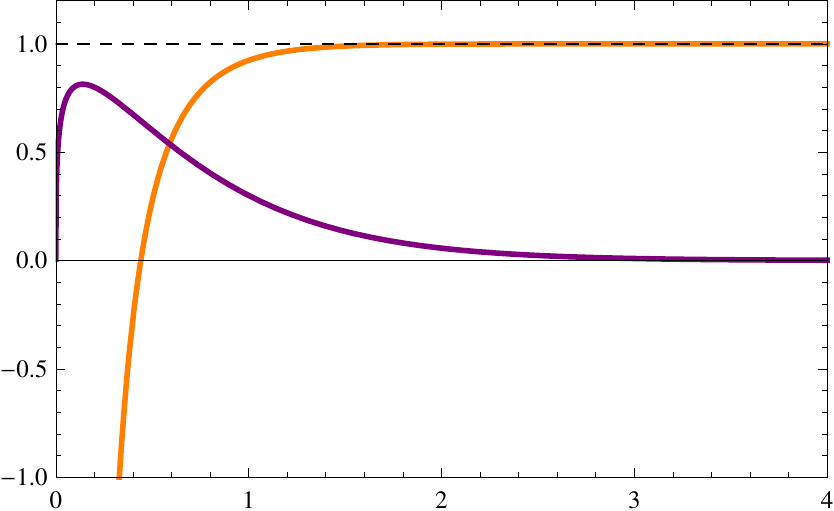}
\caption{${\cal H}^{(2,2)}$ Schr\"odinger equation potential and zero-mode $\xi_0$}
\end{figure}

Justifying the singularity of the bound state as $\rho\rightarrow0$ requires introduction of some other element into the solution. It turns out that what can be included nicely is an NS-5 brane. The asymptotic structure of the Salam-Sezgin background as $\rho\rightarrow0$ limits to the horizon structure of a NS-5 brane. This  also allows for the inclusion of an additional NS-5 brane source as $\rho\rightarrow0$. After such an inclusion, the zero-mode transverse wavefunction $\xi_0$ remains {\em unchanged}. Moreover, inclusion of such an additional NS-5 brane does not alter the 8 unbroken space-time supersymmetries possessed by the Salam-Sezgin background. The NS-5 modified $D=10$ supergravity solution can still be given explicitly for the metric, dilaton and 2-form gauge field \cite{Crampton:2014hia}:
\bea
d\hat s_{10}^2 &=&  H^{-\frac14}(dx^\mu dx_\mu + dy^2 + 
  \fft1{4 g^2}\, [d\psi + \sech 2\rho\, (d\chi+\cos\theta\, d\varphi)]^2)+ H^{\frac34}\,d\bar s^2\cr
e^{\hat\phi} &=& H^{\frac12}\,,\quad  
\hat A_2 = \fft1{4g^2}\, \Big[ (1-c_2)\, d\chi + \sech 2\rho\, d\psi\Big]
    \wedge (d\chi+ \cos\theta\, d\varphi)\,.
\eea
where now
\be
H = c_1 + c_2\, \log\tanh\rho + \sech 2\rho
\ee
where $c_1$ and $c_2$ are integration constants.

\begin{figure}[H]
\captionsetup{width=.85\linewidth}
\centering
\includegraphics[scale=.35]{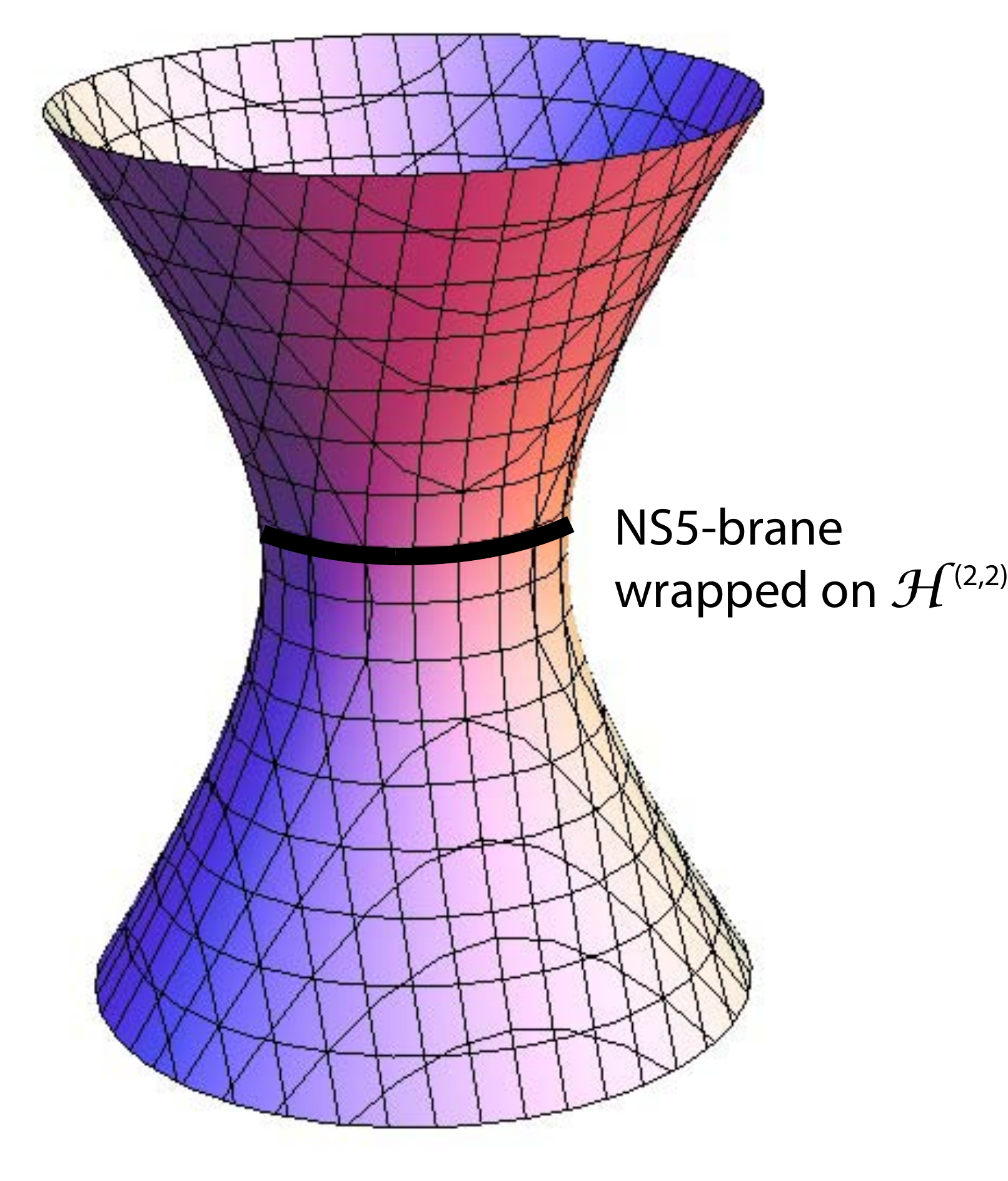}
\caption{
${\cal H}^{(2,2)}$ space with an NS-5 brane source wrapped around its `waist'
and smeared on a transverse $S^2$.}
\end{figure}

\section*{\large Braneworld Effective Gravity}

The effective action for 4D gravity reduced on the background Salam-Sezgin solution is obtained by letting the higher dimensional metric take the form
$d\hat s^2=e^{2A(z)}(\eta_{\mu\nu}+h_{\mu\nu}(x)\xi_0(\rho))dx^\mu dx^\nu + \hat g_{ab}(z)dz^adz^b$, where the warp factor $A(z)$ and the transverse metric $\hat g_{ab}(z)$ are given by the Salam-Sezgin background.

Starting from the 10D Einstein gravitational action
\be
I_{\sst{10}}={1\over 16\pi G_{\sst{10}}}\int d^{10}x \sqrt{\hat g} 
\hat R(\hat g)
\ee
and making the reduction to 4D in accordance with the previous discussion, one obtains at quadratic order in $h_{\mu\nu}$ the linearized 4D Einstein (\ie massless Fierz-Pauli) action with a prefactor $\upsilon_0^{-2}$
\begin{align}
&I_{\rm\sst{lin\,4}}=\nn\\ 
&{1\over\upsilon_0^2}\int d^4x 
\left(-\frac12\partial_\sigma h_{\mu\nu}
\partial^\sigma h^{\mu\nu}+\frac12\partial_\mu 
h^\sigma{}_\sigma\partial^\mu h^\tau{}_\tau + 
\partial^\nu h_{\mu\nu}\partial^\sigma h^\mu{}_\sigma + 
h^\sigma{}_\sigma\partial^\mu\partial^\nu h_{\mu\nu}\right)\,.
\end{align}

The normalizing factor
$\upsilon_0=
\left({16\pi G_{\sst{10}}g^5\over\pi^2\ell_yI_2}\right)^{\frac12}\nn$
involves the first of a series of integrals involving products of the transverse wavefunction $\xi_0$. For $\upsilon_0$ one needs
\be
I_2=\int_0^\infty d\rho\, \sinh2\rho\,\xi_0^2=
\frac{\pi^2}{12}\,.
\ee
 The ability to explicitly evaluate such integrals of products of transverse wave functions is directly related to the integrable-model P\"oschl-Teller structure of the transverse wavefunction Schr\"odinger equation with $V(\rho)=2-\coth^2(2\rho)$. This is reminiscent of the way in which analogous integrals for the hydrogen atom can be evaluated using the integrable structure following from its $\hbox{SO}(4)$
 symmetry \cite{Lieber:1969rg}.
 
 In order to obtain the effective 4D Newton constant, one first needs to 
rescale $h_{\mu\nu}=\upsilon_0\tilde h_{\mu\nu}$ in order to obtain a 
canonically-normalized kinetic term for $\tilde h_{\mu\nu}$. 
Then the leading effective 4D coupling $\kappa_4=\sqrt{32\pi G_{\sst4}}$ 
for gravitational self-interactions is obtained from the coefficient in front of the
trilinear terms in $\tilde h_{\mu\nu}$ in the 4D effective action. These involve the integral
\be
I_3=\int_0^\infty d\rho \sinh2\rho\,\xi_0^3=
-\fft{3\zeta(3)}{4}\,;
\ee
accordingly, the 4D Newton constant is given by
\be
G_{\sst4}={486\,\zeta(3)^2G_{\sst{10}}g^5\over\pi^8\ell_y}
\ee
with corresponding 4D expansion coupling constant
\be
\kappa_4=72\sqrt3\zeta(3)\left({G_{\sst{10}}g^5\over\pi^7\ell_y}
\right)^{\frac12}\,.
\ee

Note that the convergence of the $I_2$ and $I_3$ integrals in the evaluation of $G_{\sst4}$ is ensured by the presence of the $\sinh2\rho$ factor as $\rho\to0$ and by the asymptotic falloff of $\xi_0(\rho)$ as $\rho\to\infty$. 

By contrast, in a standard Kaluza-Klein reduction down to 4D, the transverse wavefunction would just be $\xi={\rm const}$, causing the $I_2$ integral to diverge. This would, however, give rise to a {\it vanishing} 4D Newton constant. The problem of a vanishing Newton constant after such a standard reduction on a noncompact space was pointed out early on by Hull and Warner \cite{Hull:1988jw}. As mentioned above, the standard $\xi_0={\rm constant}$ Kaluza-Klein reduction in this case yields a formally consistent truncation \cite{Cvetic:2003xr}, but the price one pays for this with an infinite transverse space is to have $G^{\sst \xi_0\,{\rm const}}_{\sst4}=0$.

The reduction with a {\em normalizable} transverse wavefunction $\xi_0(\rho)$ yields an acceptably finite $G_{\sst4}$, but at the price that the reduction does not produce a formally consistent truncation. This can be thought of as a feature rather than a bug, however, as what it means is that instead of suppressing the massive Kaluza-Klein modes, one should properly integrate them out in deriving the 4D effective theory.

The P\"oschl-Teller integrable structure of the transverse Schr\"odinger problem enables much of this to be done explicitly.
The other $\xi_0^n$ integrals needed in evaluating the leading effective theory can also be done explicitly. One finds 
\be
I_{n} \equiv 
\int_0^\infty d\rho\, \sinh2\rho\, \xi_0^n(\rho) =
(-1)^n\, n!\,2^{-n}\, \zeta(n)\,.
\ee
Moreover, integrating out the continuum of massive modes requires performing integrals like 
\be
\int_0^\infty d\rho\, \sinh2\rho\, \xi_0^n(\rho)\xi_\lambda(\rho)
\ee
which also can be evaluated and the results given in terms of Legendre functions. Integrating out the $\xi_\lambda$ contributions then produces a series of corrections to the leading-order effective theory.

The ``inconsistency'' of the reduction to $D=4$ is revealed in the types of corrections to the lower-dimensional effective theory that can arise from integrating out the massive modes. There are some similarities here to compactification on Calabi-Yau spaces \cite{Duff:1989cr}. However, in such CY compactifications, if one focuses on the parts of the leading order effective theory without scalar potentials, the result of integrating out the massive KK modes is purely to generate higher-derivative corrections to the leading order effective theory.

In the present case, however, important corrections can be obtained also in the leading order two-derivative part of the effective theory. One can see this thanks to the special integrability features of the P\"oschl-Teller transverse wavefunctions, which allow for transverse integrals actually to be done explicitly. Note, for example that quartic terms in $h_{\mu\nu}(x)$ involve the integral $I_4=4!\,2^{-4}\, \zeta(4)$. This, however, does not yet yield the expected quartic term\footnote{That the true test of a gauge invariance characteristically arises at fourth order in such an expansion has been underlined in Reference \cite{Deser:2019yig}.} with a coefficient $(\kappa_4)^2$: $I_4$ involves $\zeta(4)$, while $(\kappa_4)^2$ involves $(\zeta(3))^2$.

The deficit has to arise from the result of integrating out massive modes. The pattern is rather intricate, and involves special properties of products of the transverse wavefunctions $\xi_0$ and $\xi_\lambda$. In fact, one good way to discover such properties is to start from the 10-dimensional theory and demand that the expansion in 4-dimensional massless $h_{\mu\nu}$ modes and the massive $h^{(\lambda)}_{\mu\nu}$ modes reproduce 10-dimensional general covariance.

A study of how higher-dimensional general covariance transmits local spin-2 gauge symmetry down to the effective lower dimensional theory will be given elsewhere \cite{EHLS}. One clue to the resolution of this problem is the fact that in an appropriately constructed first-order formalism, the gravitational action can be written in a form that involves no higher than trilinear terms \cite{Deser:1969wk,Deser:2009fq}.

\section*{\large Coda}

Of Peter and his stories and his intelligence, there is much more that could be recounted. Mainly what I remember, however, is his passionate (and, at times, loud!) engagement with physics, and especially the search for fundamental theory which has engaged our community for most of our scientific lives. Peter is sorely missed, both as a pathfinding physicist and as a treasured colleague, with lifetimes of stories to tell.

\section*{\large Acknowledgment} 

This work was supported in part by the STFC under Consolidated Grant ST/P000762/1.


\begin{thebibliography}{10}

%\cite{Freund:1980xh}
\bibitem{Freund:1980xh}
  P.~G.~O.~Freund and M.~A.~Rubin,
  ``Dynamics of Dimensional Reduction,''
  Phys.\ Lett.\  {\bf 97B} (1980) 233.
  doi:10.1016/0370-2693(80)90590-0
  %%CITATION = doi:10.1016/0370-2693(80)90590-0;%%
  
%\cite{Freund:1982pg}
\bibitem{Freund:1982pg}
  P.~G.~O.~Freund,
  ``Kaluza-Klein Cosmologies,''
  Nucl.\ Phys.\ B {\bf 209} (1982) 146.
  doi:10.1016/0550-3213(82)90106-7
  %%CITATION = doi:10.1016/0550-3213(82)90106-7;%%
  
%\cite{Freund:1985je}
\bibitem{Freund:1985je}
  P.~G.~O.~Freund and P.~Oh,
  ``Cosmological Solutions With 'ten Into Four' Compactification,''
  Nucl.\ Phys.\ B {\bf 255} (1985) 688.
  doi:10.1016/0550-3213(85)90160-9
  %%CITATION = doi:10.1016/0550-3213(85)90160-9;%%
  
%\cite{Freedman:1983zt}
\bibitem{Freedman:1983zt}
  D.~Z.~Freedman, G.~W.~Gibbons and P.~C.~West,
  ``Ten Into Four Won't Go,''
  Phys.\ Lett.\  {\bf 124B} (1983) 491.
  doi:10.1016/0370-2693(83)91558-7
  %%CITATION = doi:10.1016/0370-2693(83)91558-7;%%
  
%\cite{Arafune:1974uy}
\bibitem{Arafune:1974uy}
  J.~Arafune, P.~G.~O.~Freund and C.~J.~Goebel,
  ``Topology of Higgs Fields,''
  J.\ Math.\ Phys.\  {\bf 16} (1975) 433.
  doi:10.1063/1.522518
  %%CITATION = doi:10.1063/1.522518;%%
  
%\cite{Eguchi:1976db}
\bibitem{Eguchi:1976db}
  T.~Eguchi and P.~G.~O.~Freund,
  %``Quantum Gravity and World Topology,''
  Phys.\ Rev.\ Lett.\  {\bf 37} (1976) 1251.
  doi:10.1103/PhysRevLett.37.1251
  %%CITATION = doi:10.1103/PhysRevLett.37.1251;%%
  
%\cite{Cho:1975sf}
\bibitem{Cho:1975sf}
  Y.~M.~Cho and P.~G.~O.~Freund,
  %``Nonabelian Gauge Fields in Nambu-Goldstone Fields,''
  Phys.\ Rev.\ D {\bf 12} (1975) 1711.
  doi:10.1103/PhysRevD.12.1711
  %%CITATION = doi:10.1103/PhysRevD.12.1711;%%

%\cite{Rubakov:1983bb}
\bibitem{Rubakov:1983bb}
  V.~A.~Rubakov and M.~E.~Shaposhnikov,
  ``Do we live inside a domain wall?,''
  Phys.\ Lett.\ B {\bf 125} (1983) 136.
  %%CITATION = PHLTA,B125,136;%%
  
%\cite{Randall:1999vf}
\bibitem{Randall:1999vf}
  L.~Randall and R.~Sundrum,
  ``An alternative to compactification,''
  Phys.\ Rev.\ Lett.\  {\bf 83} (1999) 4690,
  hep-th/9906064.
%%CITATION = HEP-TH/9906064;%%

%\cite{Karch:2000ct}
\bibitem{Karch:2000ct}
  A.~Karch and L.~Randall,
  ``Locally localized gravity,''
  JHEP {\bf 0105} (2001) 008,
  hep-th/0011156.
%%CITATION = HEP-TH/0011156;%%

%\cite{Bachas:2011xa}
\bibitem{Bachas:2011xa}
  C.~Bachas and J.~Estes,
  ``Spin-2 spectrum of defect theories,''
  JHEP {\bf 1106} (2011) 005,
arXiv:1103.2800 [hep-th].
%%CITATION = ARXIV:1103.2800;%%

\bibitem{Csaki:2000fc}
  C.~Csaki, J.~Erlich, T.~J.~Hollowood and Y.~Shirman,
  ``Universal aspects of gravity localized on thick branes,''
  Nucl.\ Phys.\ B {\bf 581} (2000) 309,
hep-th/0001033.
%%CITATION = HEP-TH/0001033;%%

\bibitem{Salam:1984cj}
A.~Salam and E.~Sezgin, ``{Chiral compactification on Minkowski 
$\times S^2$ of 
$N=2$ Einstein-Maxwell supergravity in six dimensions},''
\href{http://dx.doi.org/10.1016/0370-2693(84)90589-6}{{\em Phys. Lett.}
  {\bfseries B147} (1984) 47}.
%%CITATION = PHLTA,B147,47;%%.

\bibitem{Cvetic:2003xr}
M.~Cveti\v c, G.~W. Gibbons, and C.~N. Pope, ``{A string and M-theory 
origin for the Salam-Sezgin model},''
  Nucl.\ Phys.\ B {\bf 677} (2004) 164,
hep-th/0308026.
%%CITATION = HEP-TH/0308026;%%.

%\cite{Nastase:1999cb}
\bibitem{Nastase:1999cb}
  H.~Nastase, D.~Vaman and P.~van Nieuwenhuizen,
  ``Consistent nonlinear K K reduction of 11-d supergravity on AdS(7) x S(4) and selfduality in odd dimensions,''
  Phys.\ Lett.\ B {\bf 469} (1999) 96
  doi:10.1016/S0370-2693(99)01266-6
  [hep-th/9905075].
  %%CITATION = doi:10.1016/S0370-2693(99)01266-6;%%
  
\bibitem{Pugh:2010ii}
  T.~G.~Pugh, E.~Sezgin and K.~S.~Stelle,
  ``$D=7$ / $D=6$ heterotic supergravity with gauged R-symmetry,''
  JHEP {\bf 1102} (2011) 115,
arXiv:1008.0726 [hep-th].
%%CITATION = ARXIV:1008.0726;%%

%\cite{Crampton:2014hia}
\bibitem{Crampton:2014hia}
  B.~Crampton, C.~N.~Pope and K.~S.~Stelle,
  ``Braneworld localisation in hyperbolic spacetime,''
  JHEP {\bf 1412} (2014) 035
  doi:10.1007/JHEP12(2014)035
  [arXiv:1408.7072 [hep-th]].
  %%CITATION = doi:10.1007/JHEP12(2014)035;%%

\bibitem{Essin2005}
  A.~M.~Essin and D.~J.~Griffiths
  ``Quantum mechanics of the $1/x^2$ potential,''
  Am. J. Phys. {\bf 74}, 109 (2006); http://dx.doi.org/10.1119/1.2165248.
  
%\cite{Lieber:1969rg}
\bibitem{Lieber:1969rg}
  M.~Lieber,
  ``O(4) symmetry of the hydrogen atom and the lamb shift,''
  Phys.\ Rev.\  {\bf 174} (1968) 2037.
  doi:10.1103/PhysRev.174.2037
  %%CITATION = doi:10.1103/PhysRev.174.2037;%%
  
%\cite{Hull:1988jw}
\bibitem{Hull:1988jw}
  C.~M.~Hull and N.~P.~Warner,
  ``Noncompact gaugings from higher dimensions,''
  Class.\ Quant.\ Grav.\  {\bf 5} (1988) 1517.
%%CITATION = CQGRD,5,1517;%%
  
%\cite{Duff:1989cr}
\bibitem{Duff:1989cr}
  M.~J.~Duff, S.~Ferrara, C.~N.~Pope and K.~S.~Stelle,
  ``Massive {Kaluza-Klein} Modes and Effective Theories of Superstring Moduli,''
  Nucl.\ Phys.\ B {\bf 333} (1990) 783.
  doi:10.1016/0550-3213(90)90139-5
  %%CITATION = doi:10.1016/0550-3213(90)90139-5;%%
  
%\cite{Deser:2019yig}
\bibitem{Deser:2019yig}
  S.~Deser and K.~S.~Stelle,
  ``Field redefinition's help in constructing non-abelian gauge theories,''
  Phys.\ Lett.\ B {\bf 798} (2019) 135007
  doi:10.1016/j.physletb.2019.135007
  [arXiv:1908.05511 [hep-th]].
  %%CITATION = doi:10.1016/j.physletb.2019.135007;%%
  
\bibitem{EHLS}
C.~Erickson, A.~Harrold, R.~Leung and K.~S.~Stelle,
in preparation.

%\cite{Deser:1969wk}
\bibitem{Deser:1969wk}
  S.~Deser,
  ``Selfinteraction and gauge invariance,''
  Gen.\ Rel.\ Grav.\  {\bf 1} (1970) 9
  doi:10.1007/BF00759198
  [gr-qc/0411023].
  %%CITATION = doi:10.1007/BF00759198;%%
  
%\cite{Deser:2009fq}
\bibitem{Deser:2009fq}
  S.~Deser,
  ``Gravity from self-interaction redux,''
  Gen.\ Rel.\ Grav.\  {\bf 42} (2010) 641
  doi:10.1007/s10714-009-0912-9
  [arXiv:0910.2975 [gr-qc]].
  %%CITATION = doi:10.1007/s10714-009-0912-9;%%

\end{thebibliography}
\end{document}